\documentclass[11pt]{article}

\usepackage{Sweave}
\usepackage{graphicx}           
\usepackage{endnotes}            
\usepackage{amsfonts}            
\usepackage{amssymb}             
\usepackage{amsthm}              
\usepackage{amsmath}            
\usepackage[margin=1in]{geometry}
\usepackage{rotating}            
\usepackage{xspace}
\usepackage{multirow}
\usepackage{hyphenat}		
\usepackage{paralist}       
\usepackage{listings}
\usepackage{caption}
\usepackage{subcaption}
\usepackage{lmodern}
\usepackage{color}
\newcommand{\comment}[1]{} 

\usepackage{setspace}                

\usepackage{natbib}
\renewcommand{\bibsection}{\section{References}}
\bibpunct{(}{)}{;}{a}{,}{;}

\newcommand{\citepos}[1]{}
\renewcommand{\citepos}[1]{\citeauthor{#1}'s (\citeyear{#1})}

\begin{document}

\title{Yet Another Statistical Analysis of Bob Ross Paintings}

\author{Christopher Steven Marcum, PhD \\ National Institutes of Health}
\date{\today{}}
\maketitle

\abstract{
In this paper, we analyze a sample of clippings from paintings by the late artist Bob Ross. Previous work focused on the qualitative themes of his paintings \citep{hickey2014sawb}; here, we expand on that line of research by considering the colorspace and luminosity values as our data. Our results demonstrate the subtle aesthetics of the average Ross painting, the common variation shared by his paintings, and the structure of the relationships between each painting in our sample. We reveal, for the first time, renderings of the average paintings and introduce ``eigenross'' components to identify and evaluate shared variance. Additionally, all data and code are embedded in this document to encourage future research, and, in the spirit of Bob Ross, to teach others how to do so.
{\\{\linebreak \bf Keywords}: art, Bob Ross, paintings, linear subspace}
}

\doublespace

\section{Introduction}
Painter Bob Ross (1942--1995) was an icon of American art education. For over a decade, his television program, \emph{The Joy of Painting}, taught and entertained millions of Americans tuning in to the half-hour show on PBS. In the course of his art career, Ross is estimated to have painted upward of $25,000$ paintings. As a master of the Alexander ``wet-on-wet'' oil painting technique, Ross's iconic ``happy'' clouds, mountains, streams, and, of course, trees were laid down on canvas in a matter of seconds. Recently, his set of paintings became the subject of a popular blog post titled ``A Statistical Analysis of the Work of Bob Ross'' by Walt Hickey on Nate Silver's pop-stat site \href{http://fivethirtyeight.com/features/a-statistical-analysis-of-the-work-of-bob-ross/}{fivethirtyeight.com}. The post ``went viral'' on social media sites and is the inspiration for the current work. 

As a commendable digest, Hickey's approach to Ross's work should rightly be characterized as a statistical analysis of qualitative features, subjects, and themes of the paintings. He enumerates the frequency distribution of various aesthetic elements (trees, rocks, hills, etc) and, with great levity, calculates and describes the conditional probability that Ross paints one element given that he's already painted another  More, Hickey delved into the voluminous library of episodes of \emph{The Joy of Painting} to determine such statistical anomalies of the presence of humans (n=2), and chimneys (n=1), in Ross's work. Hickey also employed the k-means clustering algorithm on the data represented by these features to determine unique subsets of paintings. In this paper, we take a different approach that advances this prior work in a quantitative analysis of digital representations of Ross's paintings.

In particular, we consider a different set of research questions to be addressed by formal statistical analysis. First, what does the ``average'' Bob Ross painting look like? Here, we diverge from Hickey's approach, which describes the ``typical'' Ross painting, the features of which may be quantified using conditional probabilities as he did. Instead, we specifically want to describe average tendency in the red-green-blue colorspace of digital representations of Ross's work. That is, can we render a representation of the central tendency for Ross's work by averaging over his paintings? Second, we ask what is the common variation shared across Bob Ross's paintings? The answer to this question will shed light on the commonly held belief that Ross has a relatively standardized theme. Finally, we ask to what extent are separate Bob Ross paintings correlated with one another: what is the relationship \emph{between} Bob Ross paintings? 

Finally, this manuscript serves a didactic purpose: it illustrates how to conduct comparative quantitative research with completely reproducible results from image data packaged with the manuscript. To this end, this paper was prepared using the Sweave interface between Latex and R on a Linux operating system and the code generating the analysis is embedded in both the compiled pdf and the source. Additionally, an archive of the data can be found deposited online at the journal.

\section{Data}
The first thirty images returned from a Google image search of \linebreak ``bob+ross+paintings'' in large format (as defined by Google) and attributed to Bob Ross were downloaded on November 1st, 2014.  The selection criteria also included that the \emph{Ross} signature be present or, alternatively, that the painting could be verified against the catalog of known Bob Ross paintings from the \emph{Joy of Painting} television program which is validated by comparison with the archive on \href{http://www.tv.com/shows/the-joy-of-painting/forums/pictures-of-every-painting-15711-691600/}{tv.com}. Each image was saved using a br\%d.jp*g naming precedent, where br stands for bobross, \%d is an integer from $1$ to $30$ and $*$ is either null or the letter $e$ depending on the image source, which is used by the following embedded script.

Next, each image was cropped down to a square 550 by 550 pixels. This was automatically achieved by drawing the clipping window about the Cartesian center pixel (detected via gravity method) of each image using imagemagick and a Bourne-shell loop. The following snippet demonstrates the code, which was saved to a file called ConvertAllImages.sh. Thus, comparisons made between images are done on the pixel subset contained within this clipping window.

\begin{verbatim}
#!/bin/sh
list=`ls br*jp*`
i=1
for image in $list; do
convert $image -gravity center -crop 550X550+0+0 $i.jpg
i=$((i+1))
done
\end{verbatim}

The resulting library of clipped images can be found in Table~\ref{mat}. The intensity values of each image's three channels (red, green, and blue) was read on a pixel-by-pixel basis and stored as a three dimensional array (with dimensions $[550,550,3]$) using the ``jpeg'' library for \texttt{R}. These arrays contain the data used in the subsequent analysis. The images are sampled at 100 dpi.

\section{Analysis}

\begin{figure}
        \centering
        \begin{subfigure}[b]{0.3\textwidth}
                \includegraphics[width=\textwidth]{AvgBRdn}
                \caption{$-2 \sigma $}
                \label{fig:a1}
        \end{subfigure}
        \begin{subfigure}[b]{0.3\textwidth}
                \includegraphics[width=\textwidth]{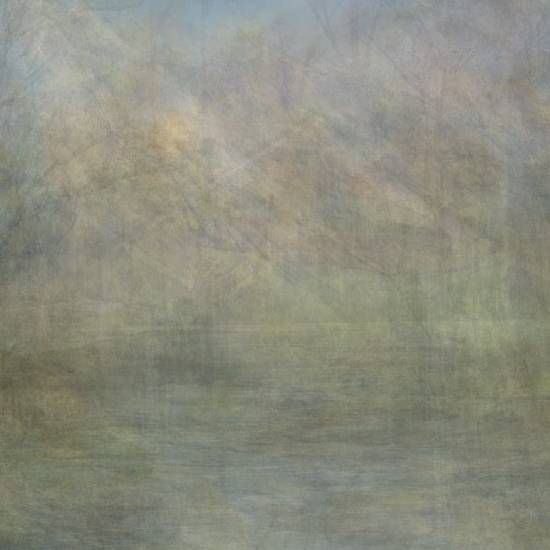}
                \caption{Average}
                \label{fig:a2}
        \end{subfigure}
        \begin{subfigure}[b]{0.3\textwidth}
                \includegraphics[width=\textwidth]{AvgBRup}
                \caption{$+2 \sigma $}
                \label{fig:a3}
        \end{subfigure}
                \begin{subfigure}[b]{\textwidth}
        \includegraphics[width=\textwidth]{AvgBR}
        \caption{Enlarged Average for Detail} 
         \label{res1a}
        \end{subfigure}
        \caption{Pixel-by-Pixel Averaging of Bob Ross Painting Central Clipping Regions (n=30) with 1.96 SDs About the Mean, All Channels Combined.}\label{res1}
\end{figure}

Our primary research question is, ``what does the average Bob Ross painting look like?'' To address this, we integrate over the respective channel indexes in the data array to obtain the mean value for each red, green, and blue channel among the 30 clippings. The resulting figure is rendered as a raster image and displayed in Figure~\ref{res1}.  Despite considerable apparent variation in the supporting set of images, this average (while quite abstract in detail) clearly shows a preference gradient for blues and pinks at the top of the image and greens and browns at the bottom. One can also detect the faint gray outlines of the trunks and branches in Ross' ``Happy Trees'' rising from the bottom toward the top of the image, suggesting that Happy Trees have low alignment variance as we'd expect. The lower and upper 95\% confidence range in these images only validates the consistency in the pixel-by-pixel averages; though, we note that the darker saturation of browns, grays, and blacks Ross used in his buildings is readily apparent in the lower bound image rendering.

Second, while the average is interesting, it fails to account for the variation in Ross' bucolic landscapes and cannot address the second research question: ``what is the common variation shared across Bob Ross' paintings?'' To examine this, we consider the eigenspace among the covariances across the dataset.  We derive a set of orthonormal vectors that best describe the shared variance across the distribution of the data---we'll call these vectors the eigenrosses. Specifically, using simple principal components analysis, we project the data back onto the eigenrosses and compare shared variances in the highly loading eigenross components. This classic approach is used in a wide variety of data reduction and statistical applications including, factor analysis, spectral analysis, and face-detection software (i.e., vis-a-vis eigenfaces methods). We conduct this for each of the three color channels as well as a flattened (monochromatic) version of the covariances---the flattened version is derived by averaging each channel with respect to each pixel, which is the Gaussian method of converting to grayscale used by most image manipulation software.

\begin{figure}
\centering
\includegraphics{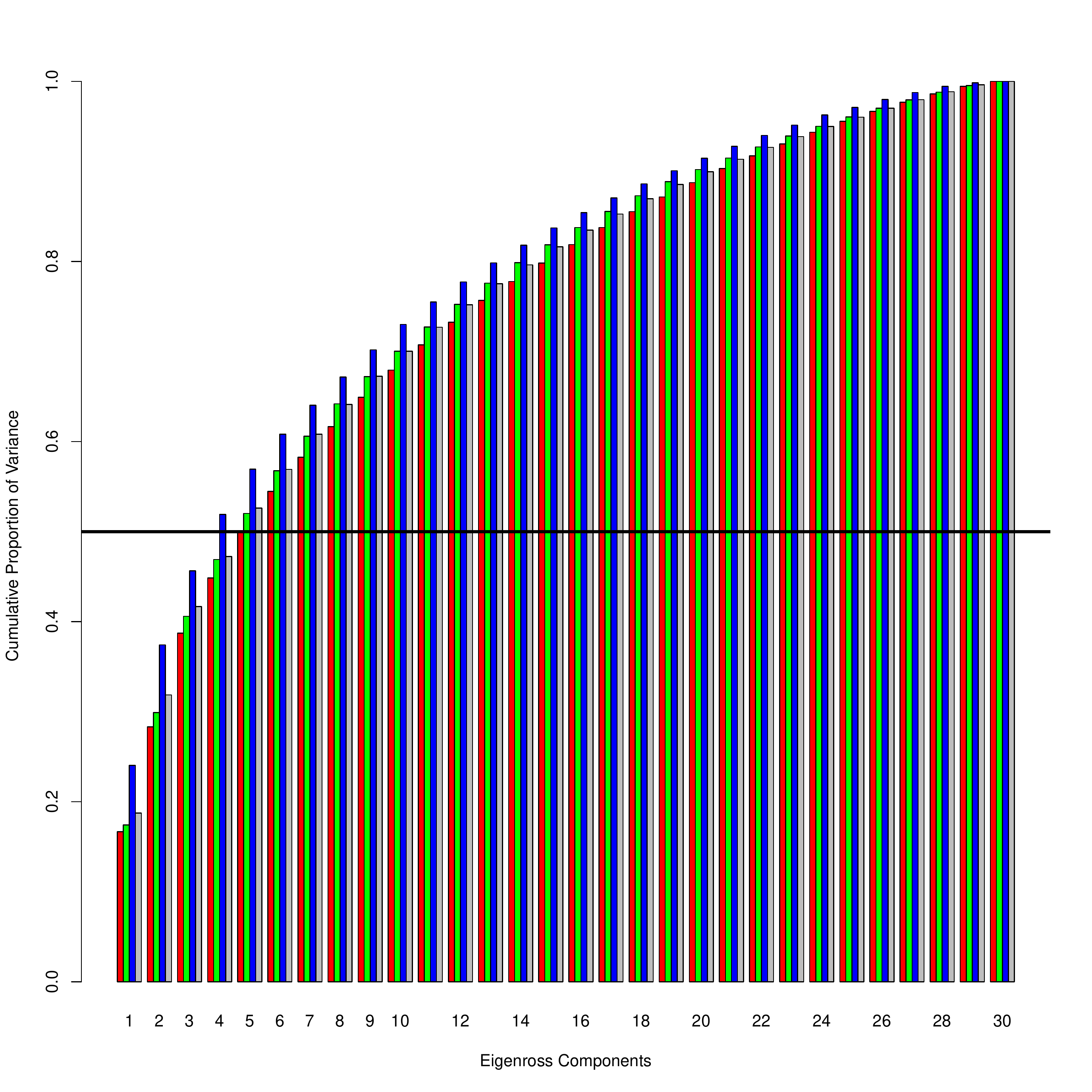}
\caption{Red-Green-Blue and Gray Channel Proportion of Shared Variances from Eigenross Components \label{res2}}
\end{figure}

The proportion of shared variances from the eigenross components is plotted in Figure~\ref{res2}. Interestingly, despite the commonly held belief that Ross' paintings are relatively similar, the plot demonstrates considerable variation---it's not until the fifth eigenross component is reached until 50\% of common variation across the whole set is attained for all channels including the flattened version (in gray). However, the first two components jointly explain more than 30\% of the variance. We can explore this further by rendering images from these components.

\begin{figure}
\centering
\includegraphics{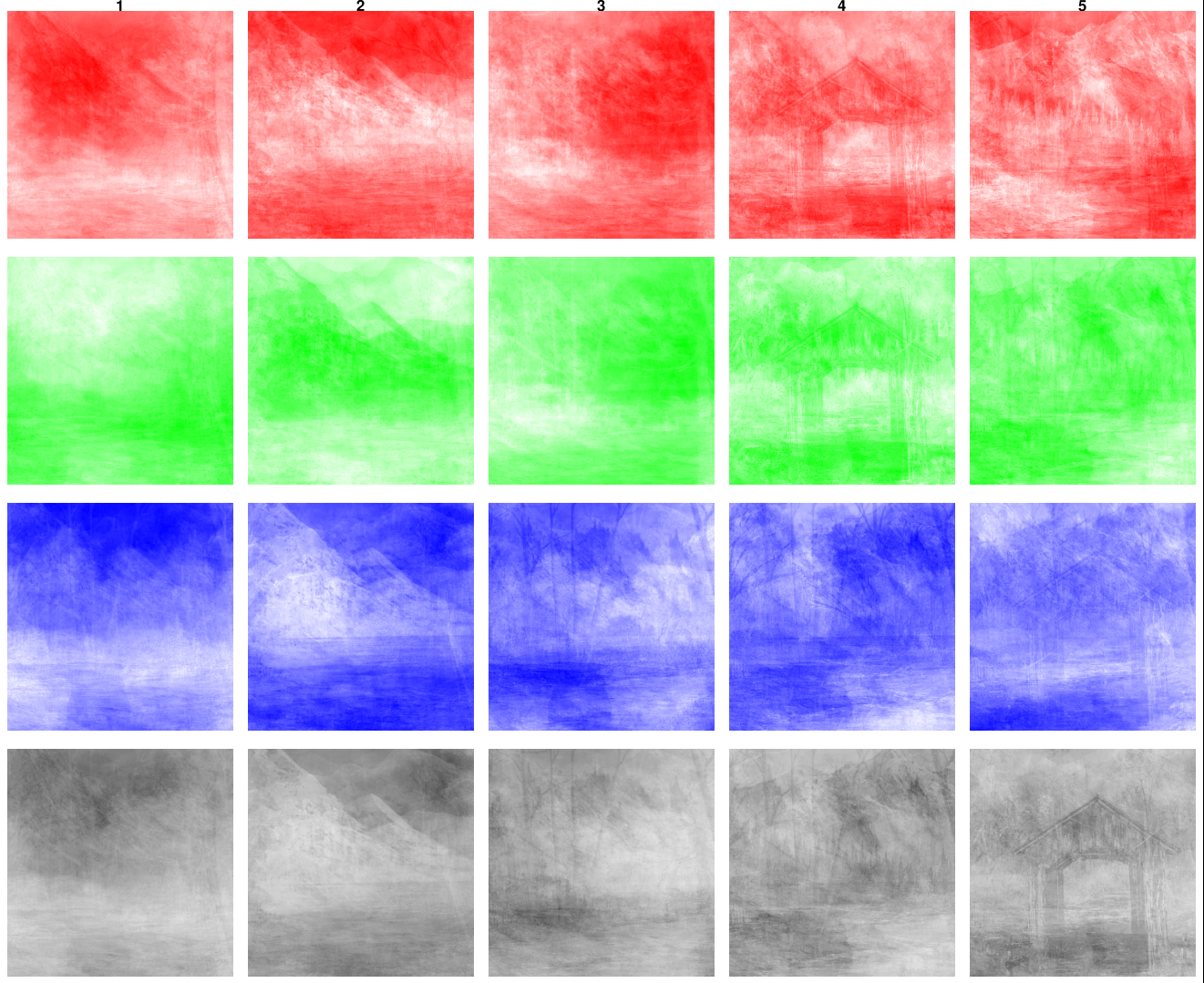}
\caption{First Five Red-Green-Blue and Gray Channel Eigenross Components \label{eigenross}}
\end{figure}

Each of the first five eigenross components is displayed in Figure~\ref{eigenross}. As these are channel-independent orthonormal transformations we cannot recombine the red, green, and blue eigenross components in a reasonable way; thus, the components are plotted separately. Lighter colored areas indicate lower pixel-by-pixel shared variance across the set of clippings in that channel. Examination of the first two components demonstrates a clear preference for upper sky and lower foreground shared variation in the red and blue channels, and a clear foreground preference in the greens. The remainder of the eigenross components appear to differentiate trees, mountains, and buildings in an order different for each channel.

\begin{figure}
\centering
\includegraphics{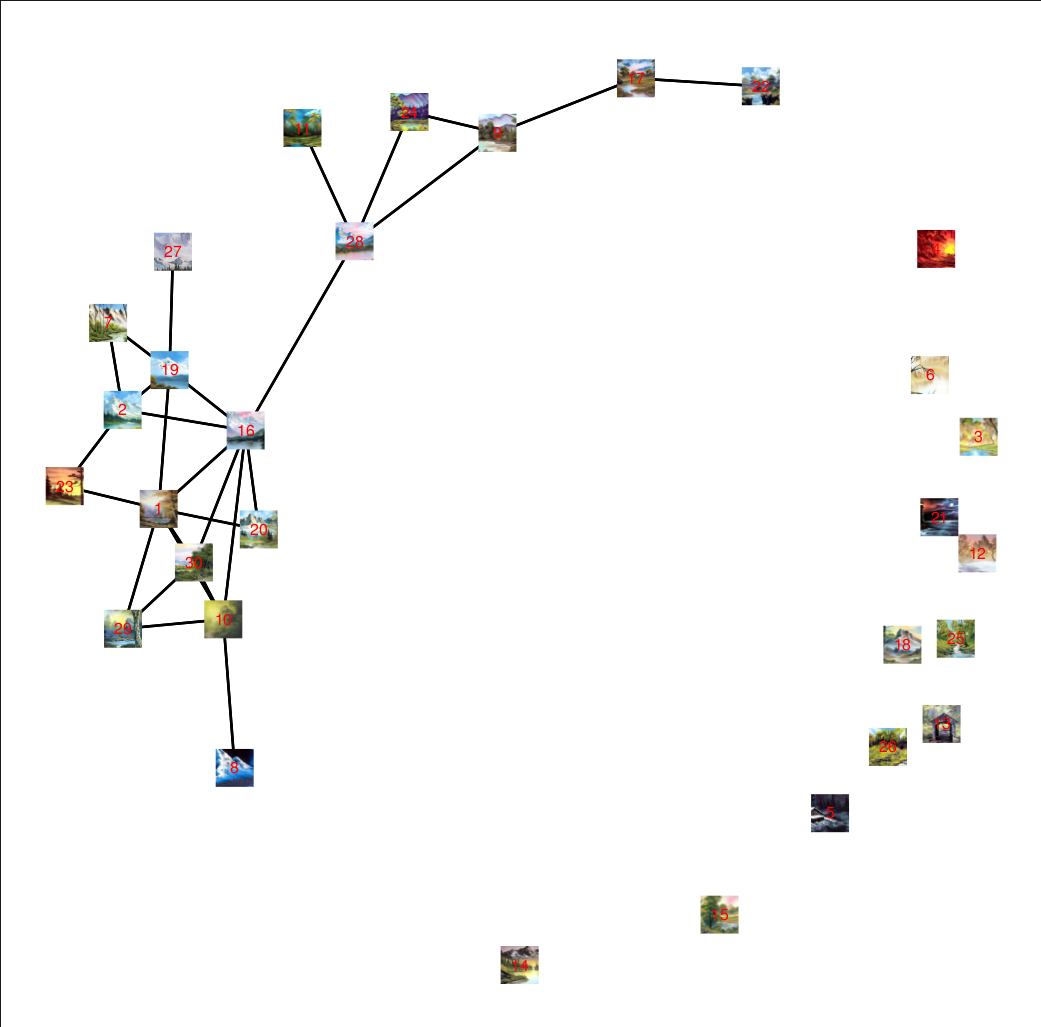}
\caption{Total Correlation Network of Bob Ross Paintings \label{network}}
\end{figure}

Finally, to address the third question, ``what is the relationship between Bob Ross paintings'' we simply examine the correlation structure using a network perspective. Specifically, we posit a relationship between two paintings if the product of red, green, and blue channel correlations is greater than or equal to $0.3^3$; in other words, two paintings are said to be related if the total correlation between them is moderate by classical standards \citep{cohen1988spab}. The resulting network is depicted in Figure~\ref{network}. 

Nearly half (n=12) of the data are isolates in this network. The isolates include the three paintings that feature buildings. In the connected component, there are two clusters, bridged by the relationship between paintings 16 and 28. Paintings, 1 and 16 appear to be the most central. These qualitative interpretations of the plot are confirmed quantitatively in Table~\ref{net}, which reports degree (number of connections), betweenness (number of non-redundant shortest-paths), and closeness (measure of being in the middle of the network) centrality scores for the paintings in the connected component of the network \citep{wasserman&faust1994snam}. 

\section{Conclusion}
As we mark the $20^{th}$ anniversary of Bob Ross's death this year, the popularity of \textit{The Joy of Painting} is again on the rise. Various public tributes have surfaced recently, including a video mash-up set to music called ``Happy Trees'' by PBS on YouTube, the selection of a Bob Ross themed costume as the winner of the Smithsonian National Zoo's Annual ``Night of the living zoo '' Costume Contest in 2013, a weekly ``Bob Ross Night'' in Missoula \url{http://www.zootownarts.org/bobross}, and now two statistical analyses (one qualitative and one quantitative) of his work. 

In this paper, we've conducted yet another statistical analysis of Bob Ross paintings. Rather than examine the qualitative features of the subjects as prior work has done, we used the quantitative values of the colorspace in digital representations of the paintings as our data. We've demonstrated the subtle aesthetics of the average Bob Ross painting, the common variance shared by a set of Ross's work, and the structure of relationships between this sample using relatively simple quantitative techniques. 

Finally, there are a number of limitations with the current approach. First, we consider only a very small sample (n=30) of the publicly available set of Ross's work, a corpus many thousands of paintings in number. Future work may wish to expand this dataset. Indeed, the techniques employed here may be used to identify specific works attributed to an artist in a larger corpus of mixed artists' work \citep{cutzu3&2005dpp}. Second, to standardize the dataset we take only the innermost central region; thus, we have a sample within a sample. It's possible that this strategy does not fully represent the variety of work encompassed by the whole lot---however, given the high spectral variability reported in our results, we believe this strategy is indeed representative of his works. Third, the source digital images that we collected from the internet were not rendered in a uniform manner; in an ideal data scenario, digital reproductions would have been obtained using the same high-resolution equipment in a controlled lighting environment. This gives rise to random errors in the channel values. However, these limitations did not impede evaluation of the research questions set forth here. As a proof-of-concept, the fact that our methods were able to recover discernible features in both the average and the variance of the set of paintings lends confidence to our results. We leave it to future research to further this approach by mitigating these limitations with a larger, more  standardized, sample. Additionally, future research should take a mixed-methods approach to statistically combine the results of qualitative and quantitative analysis of a body of art in this manner.

\bibliography{csm2}

\section{Appendix}
\begin{Schunk}
\begin{Sinput}
> require(xtable)
> require(jpeg)
> require(sna)
> require(network)
\end{Sinput}
\end{Schunk}

\begin{Schunk}
\begin{Sinput}
> system('sh ConvertAllImages.sh')
> b1<-xtable(matrix(paste("\\includegraphics[width=55]
+ {",dir(pattern="^[[:digit:]]+.jpg"),"}",sep=""),ncol=5),
+ label="mat",caption="Matrix of 550x550 pixel clipped 
+ central regions of Bob Ross Paintings. The values in the
+  red, green, and blue channels of these clippings are the
+  data used in our quantitative analysis.")
> print(b1,sanitize.text.function=identity,type="latex",
+ floating=TRUE, table.placement="h")
\end{Sinput}
\end{Schunk}

\begin{table}[ht]
\centering
\includegraphics[width=\textwidth]{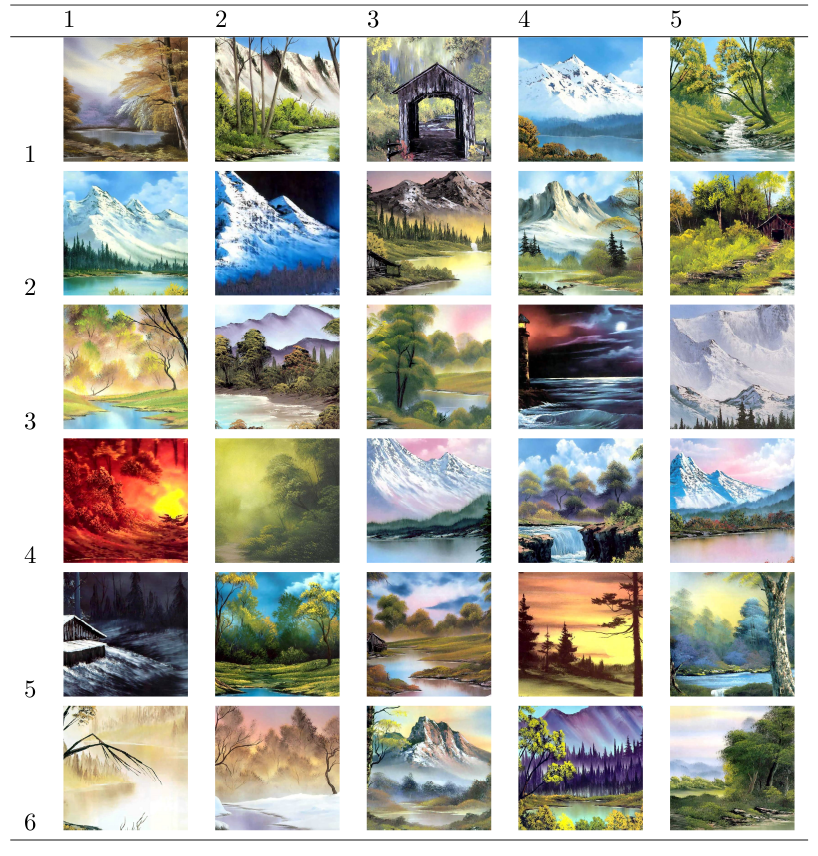}
\caption{Matrix of $550 \times 550$ pixel clipped central regions of Bob Ross Paintings. The values in the red, green, and blue channels of these clippings are the data used in our quantitative analyses \label{mat}}
\end{table}

\begin{Schunk}
\begin{Sinput}
> brfns<-dir(pattern="^[[:digit:]]+.jpg")
> datlist<-lapply(as.list(brfns),readJPEG)
> datlist.flat<-lapply(datlist,function(x) (x[,,1]+x[,,2]+
+ x[,,3])/3)
> R1<-Reduce("+",lapply(datlist,function(x) x[,,1]))/30
> G1<-Reduce("+",lapply(datlist,function(x) x[,,2]))/30
> B1<-Reduce("+",lapply(datlist,function(x) x[,,3]))/30
> writeJPEG(array(c(R1,G1,B1),dim=c(550,550,3)),target=
+ "AvgBR.jpg")
> R1.dn<-R1-(1.96*Reduce("+",lapply(datlist,function(x)
+  (x[,,1]-R1)^2))/30)
> G1.dn<-G1-(1.96*Reduce("+",lapply(datlist,function(x)
+  (x[,,2]-G1)^2))/30)
> B1.dn<-B1-(1.96*Reduce("+",lapply(datlist,function(x)
+  (x[,,3]-B1)^2))/30)
> writeJPEG(array(c(R1.dn,G1.dn,B1.dn),dim=c(550,550,3))
+ ,target="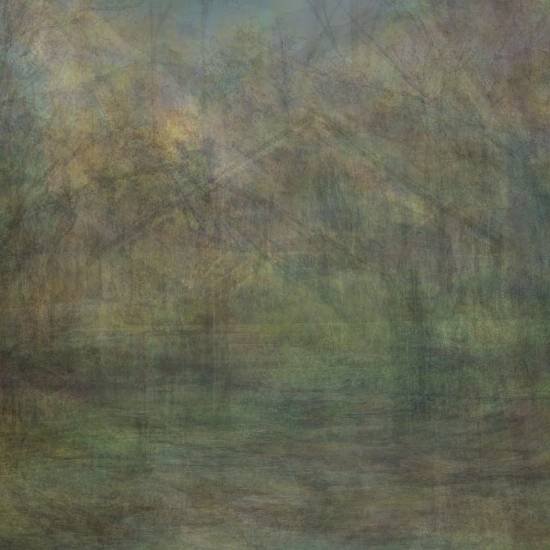")
> R1.up<-R1+(1.96*Reduce("+",lapply(datlist,function(x)
+  (x[,,1]-R1)^2))/30)
> G1.up<-G1+(1.96*Reduce("+",lapply(datlist,function(x)
+  (x[,,2]-G1)^2))/30)
> B1.up<-B1+(1.96*Reduce("+",lapply(datlist,function(x)
+  (x[,,3]-B1)^2))/30)
> writeJPEG(array(c(R1.up,G1.up,B1.up),dim=c(550,550,3))
+ ,target="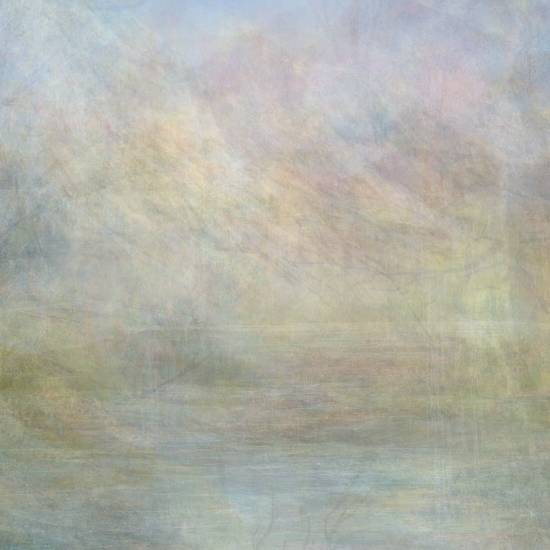")
> 
\end{Sinput}
\end{Schunk}

\begin{Schunk}
\begin{Sinput}
> R2<-prcomp(do.call("cbind",lapply(datlist,function(x)
+  c(x[,,1]))))
> G2<-prcomp(do.call("cbind",lapply(datlist,function(x)
+  c(x[,,2]))))
> B2<-prcomp(do.call("cbind",lapply(datlist,function(x)
+  c(x[,,3]))))
> flat.pc<-prcomp(do.call("cbind",lapply(datlist.flat,
+ function(x) c(x))))
> reds<-colorRampPalette(c(rgb(1,0,0),rgb(1,1,1)))
> greens<-colorRampPalette(c(rgb(0,1,0),rgb(1,1,1)))
> blues<-colorRampPalette(c(rgb(0,0,1),rgb(1,1,1)))
> fim <- function(m) t(m)[,nrow(m):1]
> pdf("eigenrosses.pdf",height=9,width=11)
> par(mfrow=c(4,5),mar=c(.5,.5,.75,.5))
> sapply(1:5, function(x)
+  image(fim(matrix(t(R2$x[,x]),nrow=550)),col=
+  reds(1000),useRaster=TRUE,axes=FALSE,main=x))
\end{Sinput}
[[1]]
NULL

[[2]]
NULL

[[3]]
NULL

[[4]]
NULL

[[5]]
NULL\begin{Sinput}
> sapply(1:5, function(x)
+  image(fim(matrix(t(G2$x[,x]),nrow=550)),col=
+  greens(1000),useRaster=TRUE,axes=FALSE))
\end{Sinput}
[[1]]
NULL

[[2]]
NULL

[[3]]
NULL

[[4]]
NULL

[[5]]
NULL\begin{Sinput}
> sapply(1:5, function(x)
+  image(fim(matrix(t(B2$x[,x]),nrow=550)),col=
+  blues(1000),useRaster=TRUE,axes=FALSE))
\end{Sinput}
[[1]]
NULL

[[2]]
NULL

[[3]]
NULL

[[4]]
NULL

[[5]]
NULL\begin{Sinput}
> sapply(1:5, function(x)
+  image(fim(matrix(t(flat.pc$x[,x]),nrow=550)),col=
+  gray.colors(1000),useRaster=TRUE,axes=FALSE))
\end{Sinput}
[[1]]
NULL

[[2]]
NULL

[[3]]
NULL

[[4]]
NULL

[[5]]
NULL\begin{Sinput}
> dev.off()
\end{Sinput}
null device 
          1 \begin{Sinput}
> pdf("cumvar.pdf",width=11,height=11)
> barplot(rbind(cumsum(prop.table(R2$sdev^2)),
+  cumsum(prop.table(G2$sdev^2)),
+  cumsum(prop.table(B2$sdev^2)),
+  cumsum(prop.table(flat.pc$sdev^2))),
+  beside=TRUE,col=c("red","green","blue","gray"),
+  names.arg=1:30,xlab="Eigenross Components",
+  ylab="Cumulative Proportion of Variance")
> abline(h=.5,col="black",lwd=3)
> dev.off()
\end{Sinput}
null device 
          1 \begin{Sinput}
> 
\end{Sinput}
\end{Schunk}

\begin{Schunk}
\begin{Sinput}
> points.with.raster<-function(x,y,raster,width=10,height=10,...){
+   x1<-x-(width/2)
+   x2<-x+(width/2)
+   y1<-y-(height/2)
+   y2<-y+(height/2)
+   points(x,y,...)
+   for(i in 1:length(x)){
+       rasterImage(raster[[i]],x1[i],y1[i],x2[i],y2[i])
+    }
+ }
> set.seed(2014)
> pdf("network.pdf",width=11,height=11)
> par(mar=c(1,1,1,1))
> totcor<-cor(do.call("cbind",lapply(datlist,function(x)
+  c(x[,,1]))))*
+  cor(do.call("cbind",lapply(datlist,function(x) c(x[,,2]))))*
+  cor(do.call("cbind",lapply(datlist,function(x) c(x[,,3]))))
> diag(totcor)<-0
> gr1<-gplot(totcor>=.3^3,usearrows=FALSE)
> points.with.raster(gr1[,1],gr1[,2],datlist,1,1)
> text(gr1[,1],gr1[,2],labels=1:nrow(gr1),col="red")
> dev.off()
\end{Sinput}
null device 
          1 \begin{Sinput}
> d1<-as.network(totcor>=.3^3,directed=FALSE)
> d1.isos<-isolates(d1)
> delete.vertices(d1,d1.isos)
> b2<-xtable(matrix(c(paste("\\includegraphics[width=25]
+ {",dir(pattern="^[[:digit:]]+.jpg")[-
+ d1.isos],"}",sep=""),round(degree(d1,gmode="graph"),4),
+ round(betweenness(d1,gmode="graph"),3),
+ round(closeness(d1,gmode="graph"),3)),
+ ncol=4, dimnames=list((1:nrow(gr1))[-
+ d1.isos],c("x","Degree","Betweenness","Closeness"))),
+ label="net",caption="Painting Centrality Statistics 
+ from Total Correlation Network")
> print(b2,sanitize.text.function=identity,type="latex",
+ floating=TRUE, table.placement="h",include.rownames=TRUE)
\end{Sinput}
\end{Schunk}

\begin{table}[h]
\centering
\includegraphics[scale=1]{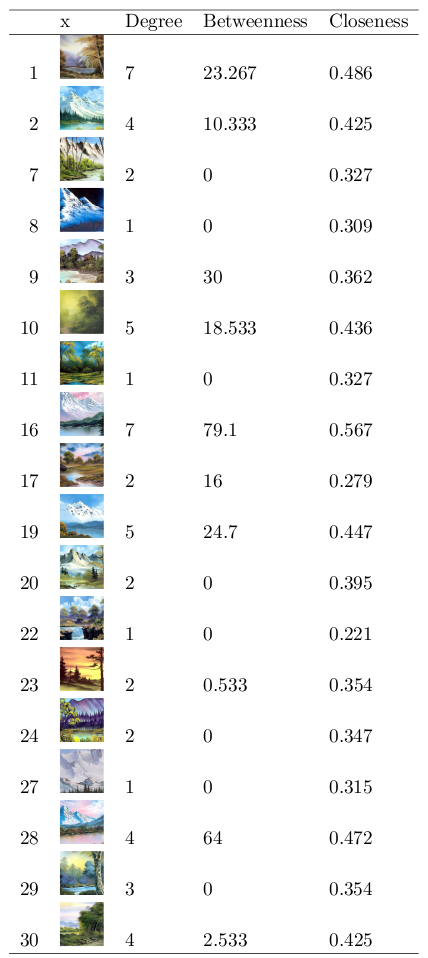}
\caption{Painting Centrality Statistics 
from Total Correlation Network \label{net}} 
\end{table}

\begin{Schunk}
\begin{Sinput}
> Stangle(file="BobRoss.Rnw",output="BobRoss.R")
\end{Sinput}
\begin{Soutput}
Writing to file BobRoss.R 
\end{Soutput}
\end{Schunk}

\end{document}